\documentclass{iaus}
\usepackage{graphicx}
\usepackage{hyperref}

\title[Morphology in Planetary Nebulae] 
{Shape, Structure, and Morphology in Planetary Nebulae}

\author[R.~A.\ Shaw]   
{Richard A. Shaw$^1$}

\affiliation{$^1$National Optical Astronomy Observatory, Tucson, AZ 85719, USA
\\email: {\tt shaw@noao.edu}
}

\pubyear{2012}
\volume{283}  
\pagerange{1--8}
\jname{Planetary Nebulae: An Eye to the Future}
\editors{A. Manchado, D. Sch{\"o}nberner, \& L. Stanghellini, eds.}

\begin{document}

\maketitle

\begin{abstract}
A revival over the past two decades in planetary nebula (PN) morphological studies springs from a combination of factors, including the advent of wide-area, high dynamic range detectors; the growing archives of high resolution images from the X-ray to the sub-mm; and the advent of sophisticated models of the co-evolution of PNe and their central stars. Yet the story of PN formation from their immediate precursors, the AGB stars, is not yet fully written. PN morphology continues to inspire, provide context for physical interpretation, and serve as an ultimate standard of comparison for many investigations in this area of astrophysics. After a brief review of the remarkable successes of PN morphology, I summarize how this tool has been employed over the last half-decade to advance our understanding of PNe. 

\end{abstract}

\firstsection 
\section{Introduction}

Since planetary nebulae (PNe) were recognized as a distinct astrophysical phenomenon nearly a one-and-a-half centuries ago, the role of morphology in PN research has evolved from simple description to a key tool in understanding their origins, structure, and evolution. Over the last century the study of morphology has provided many profound insights into the PN phenomenon. 
Morphology inspires, clarifies common features in complicated data, connects seemingly unrelated phenomena, and provides context for interpreting results. 
The power of PN morphology is its connection to most every important aspect of the phenomenon, among them nebular structure, kinematics, dynamics, abundances, space density, and evolution. 
This discussion will, after a brief review historical perspective, focus on the substantial progress in PN morphology achieved in the last half decade; a fuller perspective may be gained from excellent reviews by \cite{Kwok07}, \cite{Corradi06}, and \cite{BF02}. 

\section{A Brief History of Morphology}

\begin{figure}[th]
\begin{center}
 \includegraphics[width=4in]{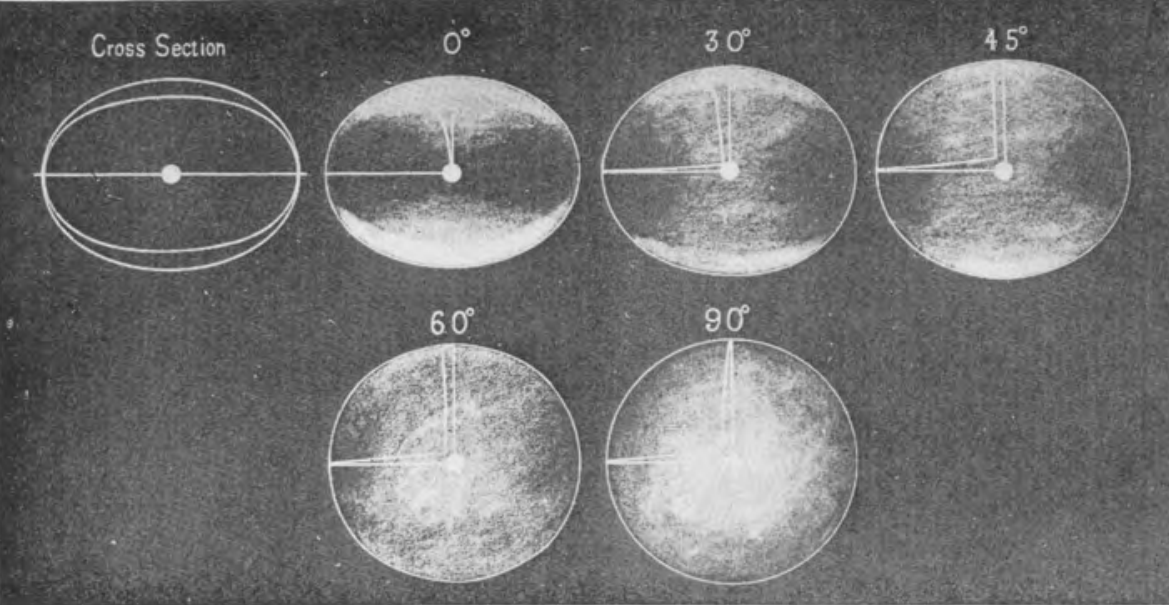} 
\end{center}
\caption{Hand drawings by \cite{Curtis18} of laboratory models showing the effect of changing perspective on inferred PN morphology. These drawings helped Curtis formulate the first comprehensive morphological classification scheme for PNe. \copyright~UC Regents/Lick Observatory; used by permission.}
\label{fig:perspective}
\end{figure}

The remarkable shapes of PNe have been noted since they were first cataloged by \cite{Messier}; hand drawings published by \cite{Secchi1877} illustrated some of the major morphological types. A seminal paper by \cite{Curtis18} established morphology as central to understanding PNe. Curtis observed: ``the details of the forms assumed by individual PNe are frequently of bewildering complexity.'' His central idea was that ``illustrations of the forms of PNe are valuable to theories of the structure and life history of planetary nebulae.'' He appreciated the impact of exposure depth and limited dynamic range in classifying PN morphology, and in many cases published hand drawings rather than his photographs to illustrate low-contrast but important detail. 
Curtis also appreciated the impact of projection on classification, and used ``semi-transparent models suspended in a liquid of nearly the same refractive index, and viewed by transmitted light'' to observe the effect of perspective on morphology. One of his hand-drawn illustrations of this effect is shown in Figure~\ref{fig:perspective}, which might well be compared to the computer models of \cite{ZhangKwok98} that were published some 80 years later. These experiments and his large collection of images of 78 northern PNe also allowed Curtis to rule out spatially variable extinction as a cause of morphological macro-structure. 

\cite{Curtis18} also noted that PNe generally have central stars, and made use of radial velocities, the number and distribution of PNe on the sky, and their distribution of angular sizes to conclude that PNe are most decidedly a rare Galactic disk phenomenon (and not related to the famous \textit{spiral nebulae} now known as external galaxies), corresponding to some brief ($\sim10^4$ yr) phase of stellar lifetimes. The combination of morphology with other information available to him at the time, along with some uncanny insights, allowed Curtis to advance the understanding of the PN phenomena to an extent seldom realized in astrophysics.  

Between the publication of the \cite{Curtis18} paper and about 1970, little emphasis on morphology was evident apart from 
deep, broad-band photographic imagery by \cite{Aller56} which illustrated very low-surface brightness nebulosity in the vicinity of many PNe. 
More quantitative morphological work was possible with the advent of monochromatic imaging in bright optical emission lines, where differences in morphology from lines of differing ionization probe structure and optical depth (\cite[see the review by Reay 1983]{Reay83}). Promising though this technique was, the application was severely limited by detector technology until the work of \cite{Balick87} and \cite{Chu87}. 
A modern revival of morphology began with the work of \cite{Greig72}, who used spatial and kinematic data to show that \textit{bi-nebulous} (now classified: bipolar) PNe originate from a younger population of stars. Interest in bipolar nebulae in particular grew with the pioneering work of \cite{Peimbert78} and \cite{PTP83}, which related elevated N and He abundances to the bipolar morphological type. 

\section{Morphology in Context}

\begin{figure}[h]
\begin{center}
 \includegraphics[width=3.8in]{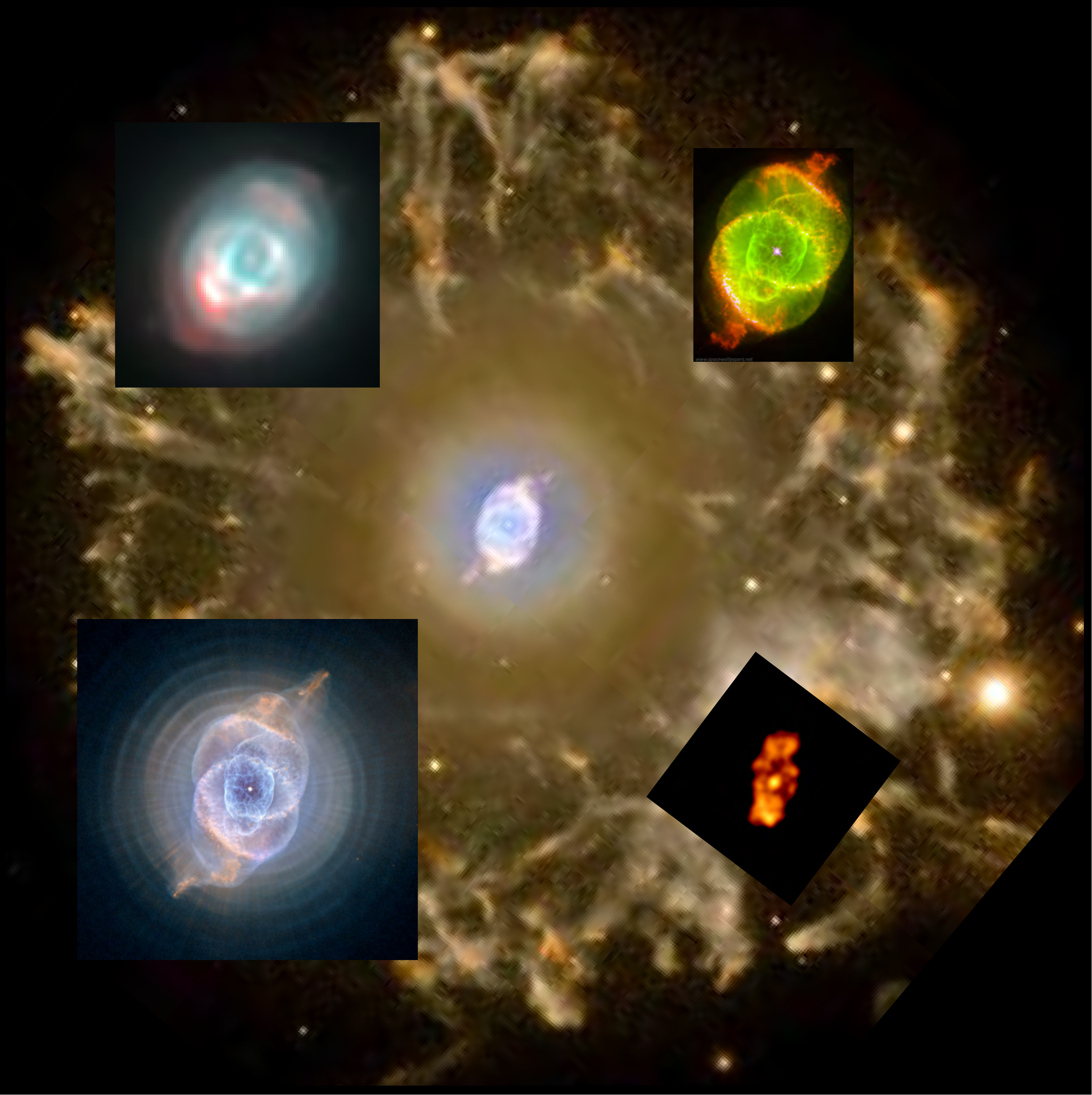} 
\end{center}
\caption{Deep, multi-band image of NGC~6543 (credit: R. Corradi/IAC) showing a large halo ({\it center}). Superimposed are four multi-band images of the nebular core: 
an ameteur ground-based image ({\it upper left}), a multi-band image from \textit{HST} ({\it upper right}; credit: J.P. Harrington and K.J. Borkowski, and NASA), a deep \textit{HST} exposure with compressed dynamic range ({\it lower left}; credit: 
NASA, ESA, HEIC) 
and an X-ray image from CXC ({\it lower right}; credit: Chu et al. and NASA). 
}
 \label{fig:catseye}
\end{figure}

Structural features can be inferred from morphology, but this must be done with care because the inferences that are drawn depend upon a variety of observational effects. Some of these factors are illustrated in multiple insets for NGC~6543 in Figure~\ref{fig:catseye}. The effect of improving angular resolution by a factor of 10 is apparent by comparing a ground-base image in the upper left to a multi-band image from \textit{HST} in the upper-right; this exposure highlighted the utility of combining multiple, narrow-band images to reveal ionization stratification and the unambiguous detection of jets emanating from the vicinity of the central star. Exposure depth and dynamic range were also important for the interpretation of concentric rings as remnant features of gas ejection on the AGB, as shown in the \textit{HST} image at lower-left. Further effects of the remnant AGB wind, and its interaction with the surrounding ISM, are clearly shown in the large, central image, while structure evident in X-rays in the lower right panel was key to understanding the role of pressure from hot gas on the dynamics of the nebula. 

Features in an image on a wide variety of spatial scales provide important morphological details; features common to many PNe form the basis of classification schemes. However, not all features are what they seem. \cite{Frew08} and \cite{FrewParker10} offer a tutorial on how to identify the genuine PNe, using all available evidence. It is vitally important to eliminate impostors from large discovery surveys such as that by \cite{MASH} using criteria including morphology, kinematics, and ionization in order to distinguish impostors from genuine PNe. 

In spite of these caveats, the greatest strides in PN morphological studies in the last few years have resulted directly from a combination of technical advances in detector technology and observing techniques, including the development of highly efficient, large format detectors; narrow-band imaging over a wide wavelength range; high resolution imaging from space-based observations; spatially resolved, high resolution spectrometry; and imaging polarimetry. These advances collectively enable researchers to identify important structural components on a variety of physical scales, connect the emission to specific atomic and molecular species with well known physical conditions (temperature, density, optical depth, etc.), infer internal PN dynamics and, importantly, build physical models of ever increasing sophistication to study the origin and evolution of individual PNe. The following sections summarize some of the recent, noteworthy accomplishments from such studies.

\section{Morphology Takes Center Stage}
\subsection{From Shape to Structure}

In morphological studies, resolution of relevant physical scales is key to interpreting form and structure correctly. While the utility of ionization maps has been apparent since the 1970s, its expert use for mapping important features has reached a new level in the study by \cite{Phillips_etal10} of NGC7009. They used narrow-band imaging from \textit{HST}, mid-IR imaging from \textit{Spitzer}, as well as space-based mid-IR spectra (from \textit{ISO} and \textit{Spitzer}), and ground-based IR and optical, long-slit spectra at a variety of positions. From these data they were able to create extinction maps; identify shock fronts, FLIERS and other features; and map important variations in electron temperature, density, and ionization that yielded a more complete picture of the relative contributions of shocks and photoionization to the total emission in this well-known PN. 

It has been recognized for decades that PNe shapes that are apparent in high-resolution images can be coupled effectively with spatially resolved, high-resolution spectra to determine the kinematics of the gas components and learn much about the true, three-dimensional structure (see the review by \cite[Corradi 2006]{Corradi06}). Such data are highly valuable for inferring the origin 
and the subsequent evolution of these structural elements 
as the PN expands into the ISM. The supporting data have historically been costly to obtain and analyze, particularly for complex and extended nebulae with low surface brightness components, but the new, on-line kinematic catalog by \cite{Lopez_etal12}\footnote{On-line catalog available at: \url{http://kincatpn.astrosen.unam.mx/}} 
should accelerate progress in this area. 
One particularly detailed study by \cite{Clark_etal10} dissected the velocity structure of NGC6751 using \textit{HST} imaging in H$\alpha$, \textit{Spitzer} mid-IR images, and ground-based, long-slit spectra at high-resolution. They modelled the 3-D morpho-kinematics using the SHAPE program (\cite[Steffen \& L{\'o}pez 2006]{Steffen_Lopez06}), which allowed them to distinguish 
a faint bipolar outflow emanating from a central ring, surrounded by a low-velocity inner halo. A kinematically distinct outer nebular halo is consistent with a remnant AGB wind that is interacting with the surrounding ambient gas. The morpho-kinematic structure provides insight into the physical processes at work in the nebula, and sets strong constraints on past gas ejection and its eventual return to the ISM. 

\subsection{Molecular and Dust Morphology}

Molecular gas has been studied in PNe for over two decades, and is known to be a major constituent of many PNe, accounting in some cases for much more mass than the photo-ionized atomic components \cite[(e.g., Kwok 2007)]{Kwok07}. However, the distribution of molecular gas in any given nebula and its impact on PN structure remained largely a mystery until the development of mid-IR imaging and mm/sub-mm interferometry with high spatial and spectral resolution. These technologies enabled \cite{Peretto_etal07} to study a massive, expanding molecular torus in the bipolar PN NGC6302 in extraordinary detail. This torus surrounds the central star in the same plane where optical emission shows a pinched waist. 
An earlier study of the dark lane in this PN by \cite{Matsuura_etal05} used imaging in monochromatic optical emission, broad-band and PAH-band mid-IR, and radio continuum to disentangle the  various, spatially overlapping structural components. These observations yielded detailed extinction maps and revealed a massive, warped dust disk that obscures part of the inner core in the optical band. The PAH emission is seen in an ionized inner shell, distinct from the dust disk, which tracks the ionizing radiation from the central star. A similar study in the emission of CO by \cite{Wang_etal08} reveals NGC2440 to have multi-polar structure, with rapid expansion in response to a wind or columnated flow from the central star. In cases such as these the existence of a confining material has been postulated based on the optical morphology, but the morphology in CO emission, coupled with the velocity information, provides a much more complete picture of the nebular structure, kinematics, and evolution. More broadly, these types of observations reinforce the point by \cite{Kwok07} that the morphology is greatly affected by details of structure and illumination, and various lines of evidence plus the morphological context are necessary to assemble a complete physical picture of PNe.  

Morphology from broad-band thermal emission can also provide the context for interpreting new features in PNe, as shown by \cite{Su_etal07}. Their mid-IR images of the highly evolved PN NGC7293 revealed a small, concentration of dust which is interpreted as a disk of debris that arises from ongoing collisions of Kuiper-Belt-like objects or comets that survived AGB evolution, and which are now illuminated by the fading central star. 

\subsection{Abundances and Morphology}

The pioneering work begun by \cite{Peimbert78},  \cite{PTP83}, and \cite{TPP97} relating N and He abundances to PN morphological type was extended to the Magellanic Clouds in a series of papers beginning with \cite{Stang_etal00}. \cite[Shaw et al. (2006, 2010, 2011)]{Shaw_etal06, Shaw_etal10, Shaw_etal11} summarized the strong correlation of macro-morphology with N abundance in the Magellanic Clouds, and \cite{Stang_etal09} showed the similar tight connection of C/O abundances to morphology. 
Of more recent interest, \cite{Stang_etal07} showed the remarkably tight connection between dust mineralogy, gas-phase abundances, and morphology for a sample of Magellanic Cloud PNe. Specifically, this sample shows a tight correlation of Carbon-rich dust features as seen in mid-IR \textit{Spitzer}/IRS spectra and C enrichment in the atomic gas. These nebulae are well segregated from Type-1 nebulae with O-rich dust, where N is greatly elevated in the gas phase. \cite{Stang_etal12} extended their mid-IR spectral survey to Galatic PNe, where the abundances of the progenitors is on average higher but more heterogeneous. In their sample of 150 targets, of which about one-third have been imaged with \textit{HST}/WFC3, a third population of mixed-chemistry dust is significant, and seems to correspond to very complicated morphologies. These and other correlations with morphological type suggest that a number of factors enter in to the formation and evolution of PNe at low metallicity, including the quantity and chemical make-up of the dust at the end of the AGB phase, and the metallicity of the progenitor. 

\section{The Origin of PN Morphologies}

PN morphological studies have for decades been focussed on understanding both the dynamics of PN structures and the origin of macrostructures, with the aim of constraining physical models of their formation and early evolution from their predecessors, the AGB stars. The recent observational emphasis has been two-fold: characterizing the distribution of morphologies of large samples of (often) mature PNe to understand the incidence of different macrostructures 
(\cite[e.g., Manchado et al. 1996, Shaw, et al. 2006; Parker et al. 2006]{Manchado_etal96, Shaw_etal06, Parker_etal06}), 
and on the other hand studying extensive samples of \textit{pre-planetary nebulae} (pre-PN: those where the central, post-AGB envelope has not yet been ionized; sometimes inappropriately termed \textit{proto-planetary nebulae}) and compact, presumably young PNe where photoionization is not yet complete. \cite{Sahai_etal11} point out that observing young and pre-PNe offers the chance of observing structure that is imposed shortly after formation on the AGB, before it is altered significantly by the rapidly evolving hot central star and its fast wind. 

Imaging surveys in broad-band optical and the near-IR of post-AGB objects over more than a decade have been key to progress in understanding the earliest phase of PN evolution 
(\cite[e.g., Ueta, et al.\ 2000, Siodmiak et al.\ 2008]{Ueta_etal00, Siodmiak_etal08}). The distribution of morphological types in these studies contrasts somewhat with surveys of Galactic PNe in that pre-PN samples tend to show a higher fraction of bi-polar types. Whether the difference in these distributions is significant in the absence of a volume-complete control sample is debatable, but the question is central to understanding how these structures originate and evolve in time. 
A remarkable imaging polarimetric study using \textit{HST}/NICMOS of pre-PN probed the structure of the circumstellar envelope (CSE) via dust-scattered, linearly polarized light; this technique reveals structure and mass-loss history even in the near vicinity of the bright central star. There are two main categories of macro-symmetry in CSEs: elliptical, and bipolar. 
\cite{Ueta_etal05} showed that the ellipticals feature an optically thin, hollow spheroidal shell and equatorial density enhancement, and \cite{Ueta_etal07} found that bipolars show a point-symmetric symmetry in polarized light attributed to the presence of an azimuthal density gradient in the CSE which is reversed from above to below the equatorial plane. The macro-morphological features in these CSEs demonstrate that the broad characteristics of PN structure are already in place just after the AGB phase.  

\begin{figure}[ht]
\begin{center}
 \includegraphics[width=4.5in]{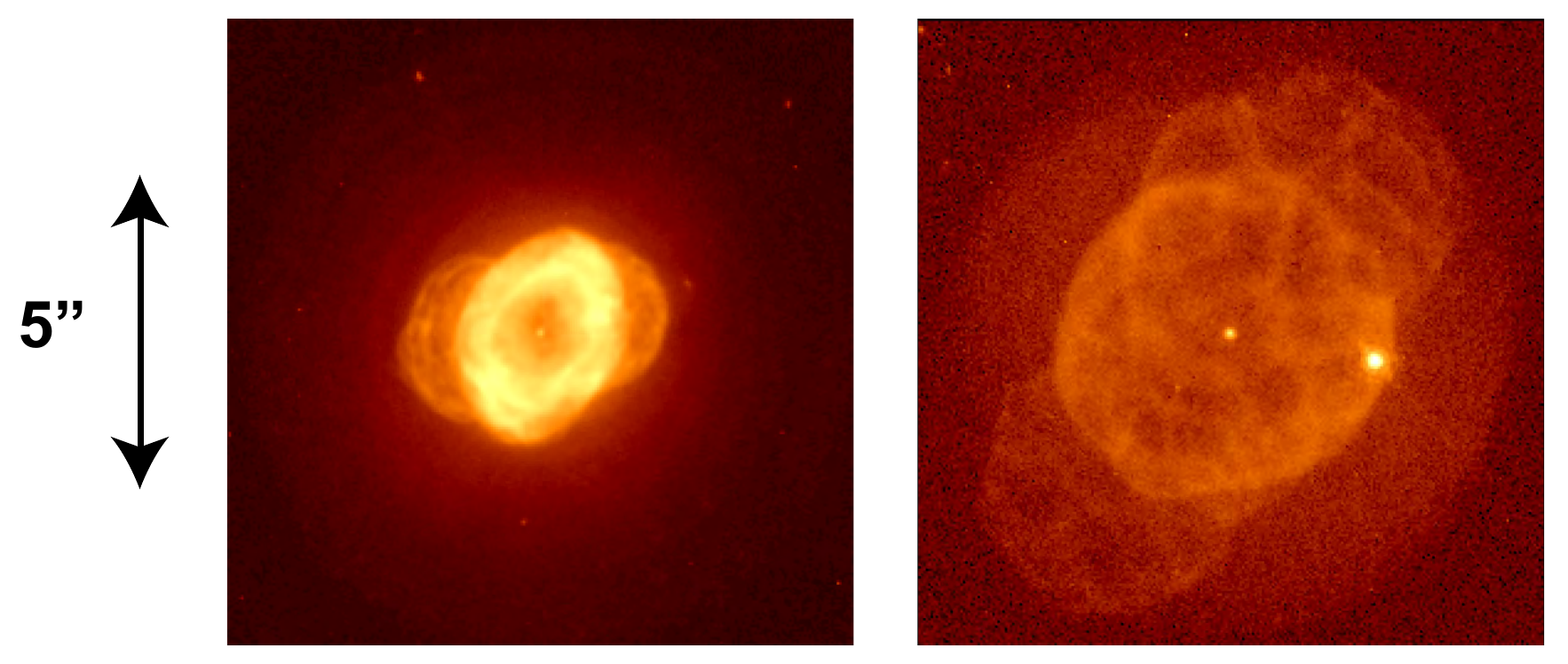} 
\end{center}
\caption{\textit{HST}/WFC3 Images of two compact Galactic PNe with similar morphologies as seen in the light of [O~{\sc iii}]: Vy1-2 (left) and VV3-5 (\textit{right}), shown in a log intensity stretch. Both appear to have central rings or tori from which a bipolar flow emerges, and which is surrounded by a faint, extended halo.
}
\label{fig:PNG}
\end{figure}

Recent work on compact Galactic PNe is begining to shed light on the early phases of PN dynamics. Such PNe, which for the most part are physically small and presumably young, are important for understanding the origin of PN morphology, just after the central star is capable of ionizing the surrounding gas. 
Large samples of high-resolution images from \textit{HST} have become available in the last decade (see the compilation by \cite[Sahai et al. 2011]{Sahai_etal11}) which shows a broad range of morphological types. Fifty of the most compact PNe (selected for published optical diameters $<5"$) have recently been observed by Shaw and collaborators with the \textit{HST}/WFC3 camera, two of which are shown in Figure~\ref{fig:PNG}. The PNe in this sample generally exhibit the full range of morphological features seen in larger Galactic PNe, although most are expected to have dynamical ages less than 2000~yr. Early indications from this new sample suggest that the effects of photoionization, heating, and a fast wind from the central star take place very quickly once it becomes hot enough to ionize the PN. 

\section{Summary}

Morphology has historically been useful for understanding the origin and evolution of PNe, and has now evolved into a critical analytical tool. Fed by advances in detector and instrument technology, space-based observing platforms, and advanced physical modelling, morphology has provided powerful insight and perspective on the origin and evolution of PNe. Modern observations are, in the context of morphology, able to identify critical components of nebular structure at all stages of PN lifetimes, view active physical processes, and therefore gather a detailed physical picture of these systems. 
As the understanding of nebular structure has deepened, new and elaborate morphological classification schemes for projected morphologies have blossomed (\cite[Sahai et al.\ 2007, 2011]{Sahai_etal07, Sahai_etal11}) which, while their applicability remains to be seen, attempt to capture the current appreciation of the connections between micro- and macro-structures at all stages of PN evolution. But perhaps the best measure of the future success of morphology in understanding PNe will be its continuing evolution as a key tool in detailed physical interpretation.

\end{document}